\documentclass{article} 
\usepackage{iclr2025_delta,times}

\usepackage{hyperref}
\usepackage{url}

\usepackage{amsthm}
\usepackage{amsmath,amsfonts,bm}

\newcommand{\bmsymbol}[1]{\ifcat\noexpand#1\relax \boldsymbol{#1} \else \mathbf{#1} \fi}

\usepackage{algorithm}
\usepackage{algorithmic}
\usepackage{graphicx}
\usepackage{subcaption}
\usepackage{booktabs}

\theoremstyle{plain}
\newtheorem{definition}{Definition}
\theoremstyle{plain}
\newtheorem{remark}{Remark}
\theoremstyle{plain}
\newtheorem{theorem}{Theorem}
\theoremstyle{plain}

\title{Trustworthy Image Super-Resolution \\ via Generative Pseudoinverse}

\author{Andreas Floros\textsuperscript{\rm $\star$}, Seyed-Mohsen Moosavi-Dezfooli\textsuperscript{\rm \textdaggerdbl} \& Pier Luigi Dragotti\textsuperscript{\rm $\star$} \\
\textsuperscript{$\star$}Imperial College London, \textsuperscript{\rm \textdaggerdbl}Apple \\
\texttt{\{andreas.floros18, pld\}@imperial.ac.uk, smoosavi@apple.com}
}

\iclrfinalcopy 
\begin{document}

\maketitle

\begin{abstract}
We consider the problem of trustworthy image restoration, taking the form of constrained sampling from the prior density. To this end, we develop generative models for the task of image super-resolution that respect the degradation process and that can be made asymptotically consistent with the low-resolution measurements, outperforming existing methods by a large margin in that respect.
\end{abstract}

\section{Introduction}

Image Super-Resolution (SR) consists of ill-posed inverse problems where the task is to recover a high-resolution image, $\bmsymbol{x}$, from Low-Resolution (LR) measurements, $\bmsymbol{y}$, which are obtained from a degradation process described by some function $\mathcal{D}:\mathcal{X}\to\mathcal{Y}$. Generative technologies are becoming increasingly more common for solving such restoration tasks. In this context, a natural concern that arises is regarding the trustworthiness of these systems. We ask the following question:

\textit{How can we maximize the quality of restored images and simultaneously minimize hallucinations?}

We take a step toward answering the above, first formalizing the notion of consistency with respect to the measurements, which is a necessary condition for trustworthiness, and then designing an algorithm with asymptotic consistency in mind. Specifically, we develop normalizing flows that allow exploration of the data manifold while ensuring consistency. We then employ latent diffusion models for this purpose, to conditionally sample missing details and recover high-quality images that are simultaneously faithful to the originals under the degradation model. Chiefly, our approach differs from existing flow-based \citep{ardizzone2018analyzing, hcflow, floros2022beyond} and diffusion-based \citep{sr3, idm, diwa} methods in two key ways:

\begin{itemize}
    \item We decouple the SR task, initially modeling the physics of the degradation function and then focusing on optimizing for restoration performance as a separate matter. This construction enables tighter control over consistency errors and, simultaneously, it allows our generative systems to solely focus on synthesizing the missing information.
    \item We recognize the strengths and weaknesses of normalizing flows and diffusion models, combining them in a way such that they complement each other and the restoration task is carried out in a structured and disentangled latent space. In doing so, we pair the interpretability of flows with the generative capacity of diffusion to tackle SR more effectively.

\end{itemize}

\section{Normalizing Flows}
\label{subsec:flows}

Flow-based models are bijections whose Jacobian determinants are typically easy to compute or approximate. Such constructions allow for density estimation and variational inference tasks via the change-of-variables formula. Of note is the affine coupling architecture \citep{realnvp}, which first partitions the data space, $\mathcal{X}$, into two parts, $\mathcal{U}$ and $\mathcal{V}$, and applies element-wise and affine invertible transformations in an alternating fashion. That is, for arbitrary $\bmsymbol{\phi}, \bmsymbol{\psi}: \mathcal{V}\to\mathcal{U}$ and $\bmsymbol{\eta}, \bmsymbol{\rho}: \mathcal{U}\to\mathcal{V}$, a single layer takes the form:
\begin{subequations}
\begin{align}
\bmsymbol{u} &\leftrightarrow [\exp\circ\bmsymbol{\psi}](\bmsymbol{v})\bmsymbol{u}+\bmsymbol{\phi}(\bmsymbol{v}), \\ \bmsymbol{v} &\leftrightarrow [\exp\circ\bmsymbol{\rho}](\bmsymbol{u})\bmsymbol{v}+\bmsymbol{\eta}(\bmsymbol{u}).
\end{align}
\end{subequations}
\begin{minipage}[t]{0.48\textwidth}
\begin{figure}[H]
    \centering
{{\includegraphics[width=\textwidth]{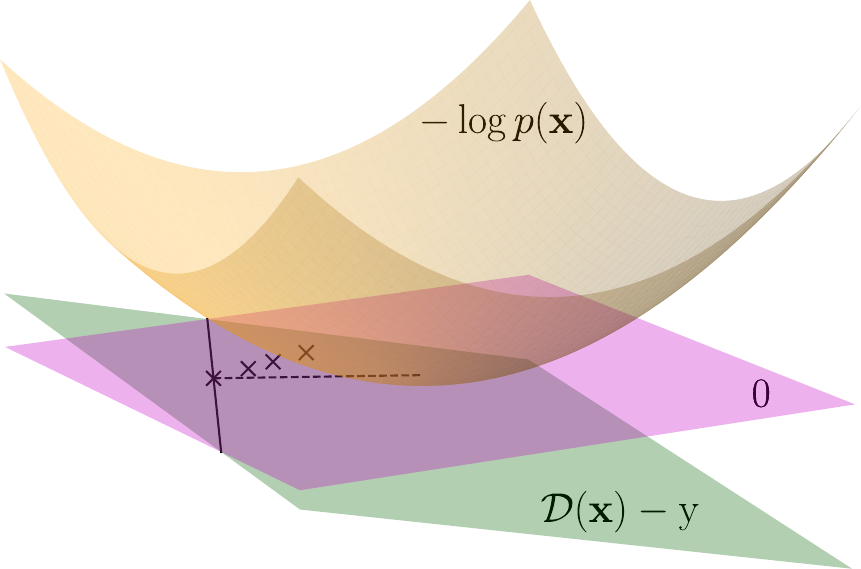} }}
    \caption{Linear SR problem in $\mathbb{R}^2$. We have $\bmsymbol{x}\sim\mathcal{N}(\bmsymbol{0},\bmsymbol{I})$ and $p(\mathrm{y}|\bmsymbol{x})=\delta_{\mathcal{D}\bmsymbol{x}-\mathrm{y}}$. The intersection of the planes $\mathrm{z}=\mathcal{D}(\bmsymbol{x})-\mathrm{y}$ and $\mathrm{z}=0$ defines the feasibility set. The marked points outside of this set correspond to solutions that optimize the prior at the cost of consistency. We deem these to not be trustworthy as they violate the constraints of the problem.}
   \label{fig:moore-penrose}
\end{figure}
\end{minipage}
\hfill
\begin{minipage}[t]{0.5\textwidth}
\begin{figure}[H]
    \centering

\subfloat{{\includegraphics[width=0.32\textwidth]{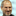} }}
    \subfloat{{\includegraphics[width=0.32\textwidth]{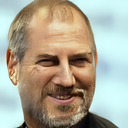} }}
    \subfloat{{\includegraphics[width=0.32\textwidth]{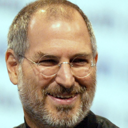} }} \\

\subfloat{{\includegraphics[width=0.32\textwidth]{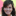} }}
    \subfloat{{\includegraphics[width=0.32\textwidth]{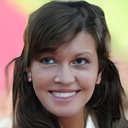} }}
    \subfloat{{\includegraphics[width=0.32\textwidth]{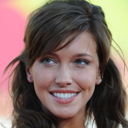} }} \\

    \setcounter{subfigure}{0}

    \subfloat[LR]{{\includegraphics[width=0.32\textwidth]{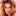} }}
    \subfloat[SR]{{\includegraphics[width=0.32\textwidth]{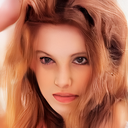} }}
    \subfloat[Original]{{\includegraphics[width=0.32\textwidth]{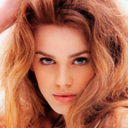} }}
    \caption{Results of our $16\times16\to128\times128$ SR.}
    \label{fig:facesrexample}
\end{figure}
\end{minipage}
\section{Diffusion Models}
Diffusion is a process by which a Gaussian distribution is gradually morphed into the desired data distribution $p$. A popular formulation is given by Denoising Diffusion Probabilistic Models (DDPMs) \citep{ddpm}, which reverse a stochastic process described by the following law:
\begin{equation}
    \label{eq:xtxt-1}p_{t|t-1}(\bmsymbol{x}_{t}|\bmsymbol{x}_{t-1})=\mathcal{N}(\bmsymbol{x}_{t};\sqrt{1-\beta_t}\bmsymbol{x}_{t-1},\beta_t\bmsymbol{I}),
\end{equation}
where $t\in[1, T]$, $\bmsymbol{x}_0\sim p$ and $\beta_t$ is the noise schedule. Reversing such a process is possible via a Gaussian denoiser that is optimal in the Mean Squared Error (MSE) sense \citep{reversediffusion, tweedie}. In practice, one trains a single noise-predicting neural network, $\bmsymbol{\epsilon}_{\bmsymbol{\theta}}(\cdot, t)$, over all time-steps $t$. Notably, the structure of Equation \ref{eq:xtxt-1} yields an efficient method for sampling $\bmsymbol{x}_t|\bmsymbol{x}_0$, avoiding multi-step simulations. Letting $\alpha_t=1-\beta_t$ and $\bar\alpha_t=\prod_{s=1}^t\alpha_s$, this allows training with a weighted Evidence Lower Bound (ELBO), which is equivalent to the following optimization \citep{ddpm}:
\begin{equation}
\min_{\bmsymbol{\theta}}\mathbb{E}_{t,\bmsymbol{x}_0,\bmsymbol{\epsilon}}\lambda_t\|\bmsymbol{\epsilon}_{\bmsymbol{\theta}}(\sqrt{\bar\alpha_t}\bmsymbol{x}_0+\sqrt{1-\bar\alpha_t}\bmsymbol{\epsilon},t)-\bmsymbol{\epsilon}\|^2_2.
\end{equation} 
\section{Proposed Method}
\label{sec:proposed}
Let $\delta$ denote the Dirac delta and let $\mathcal{D}:\mathcal{X}\to\mathcal{Y}$ be the degradation associated with our SR problem. We are interested in sampling from the posterior, $p(\cdot|\bmsymbol{y})$, given evidence, $p(\bmsymbol{y}|\bmsymbol{x})=\delta_{\mathcal{D}(\bmsymbol{x})-\bmsymbol{y}}$, and a finite data budget for approximating the prior density $p(\bmsymbol{x})$. In practice, black-box learning systems risk violation of the evidence constraints. Contrary to such approaches, we instead endeavor to design SR systems in a physics-aware manner, such that they respect consistency, defined as follows.
\begin{definition}[Consistency]
\label{def:consistency}For $\bmsymbol{y}=\mathcal{D}(\bmsymbol{x})$, we say that $\bmsymbol{x}'$ is consistent or feasible when $\mathcal{D}(\bmsymbol{x}')=\bmsymbol{y}$.
\end{definition}
The solvers we are looking for naturally take the form of generalized inverses, extending the notion from the linear case to potentially non-linear operators. Thus, we will parameterize our neural networks as such. We give our definition of generalized inverses and a motivating example below.
\begin{definition}[Generalized Inverse]\label{def:ginv}
A (stochastic) reflexive generalized inverse for $\mathcal{D}:\mathcal{X}\to\mathcal{Y}$ is any operator, $\mathcal{D}^\dagger:\mathcal{Y}\to\mathcal{X}$, satisfying $\mathcal{D}\circ\mathcal{D}^\dagger\circ\mathcal{D}\overset{d}=\mathcal{D}$ and $\mathcal{D}^\dagger\circ\mathcal{D}\circ\mathcal{D}^\dagger\overset{d}=\mathcal{D}^\dagger$ (reflexive).
\end{definition}
\begin{remark}
Consider linear SR problems under a standard Gaussian prior, detailed in Figure \ref{fig:moore-penrose}. The maximum a posteriori, consistent solution is obtained by the Moore--Penrose pseudoinverse.
\end{remark}
\begin{figure*}[!t]
    \centering
    \subfloat[LR]{{\includegraphics[width=0.16\textwidth]{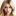} }}
\subfloat[SR3]{{\includegraphics[width=0.16\textwidth]{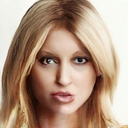} }}
\subfloat[IDM]{{\includegraphics[width=0.16\textwidth]{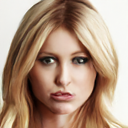} }}
\subfloat[DiWa]{{\includegraphics[width=0.16\textwidth]{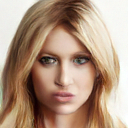} }}
\subfloat[Ours]{{\includegraphics[width=0.16\textwidth]{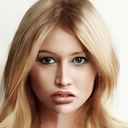} }}
\subfloat[Original]{{\includegraphics[width=0.16\textwidth]{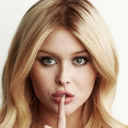} }}
    \caption{$8\times$ SR. Our method better preserves several visible facial features in this example.}
   \label{fig:facesr}
\end{figure*}
\begin{minipage}[!t]{0.548\textwidth}
\begin{algorithm}[H]
\caption{Degradation flow training\vphantom{Refinement via diffusion}}
\label{alg:learnflow}
\textbf{Input}: Data, $p(\bmsymbol{x})$, degradation function $\mathcal{D}$ \vphantom{Data, $p(\bmsymbol{x})$, trained flow $\bmsymbol{f}_{\bmsymbol{\xi}}$}\\
\textbf{Parameters}: Initial flow, $\bmsymbol{f}_{\bmsymbol{\xi}}$, $\sigma>0$ \vphantom{Initial DDPM, $\bmsymbol{\epsilon}_{\bmsymbol{\theta}}$, $\{\beta_t\}_{t=1}^T$}\\
\textbf{Output}: Learned $\bmsymbol{\xi}$
\begin{algorithmic}[1] 
\REPEAT
\STATE $\bmsymbol{x}\sim p$ \vphantom{$[\bmsymbol{y}, \bmsymbol{z}]\sim (\bmsymbol{f}_{\bmsymbol{\xi}})_{\#}p$}
\STATE $[\bmsymbol{y}, \bmsymbol{z}]=\bmsymbol{f}_{\bmsymbol{\xi}}(\bmsymbol{x})$ \vphantom{$t\sim\text{Unif}(\{1,\dots,T\})$, $\bmsymbol{\epsilon}\sim\mathcal{N}(\bmsymbol{0},\bmsymbol{I})$}
\STATE $\log p_{\mathcal{Y}\times\mathcal{Z}}(\bmsymbol{y},\bmsymbol{z})\approx-\frac{1}{2\sigma^2}\|\bmsymbol{y}-\mathcal{D}(\bmsymbol{x})\|^2_2-\frac{1}{2}\|\bmsymbol{z}\|^2_2$
\STATE Take gradient ascent step with \\ \centering $\nabla_{\bmsymbol{\xi}}\left\{\log p_{\mathcal{Y}\times\mathcal{Z}}(\bmsymbol{y}, \bmsymbol{z})+\log\left|\det\frac{\partial\left[\bmsymbol{y}, \bmsymbol{z}\right]}{\partial{\bmsymbol{x}}}\right|\right\}$ \vphantom{$\nabla_{\bmsymbol{\theta}}\|\bmsymbol{\epsilon}_{\bmsymbol{\theta}}\left(\bmsymbol{z}_t,\bmsymbol{y},t\right)-\bmsymbol{\epsilon}\|^2_2$}
\UNTIL{converged}
\RETURN{$\bmsymbol{\xi}$\vphantom{$\theta$}}
\end{algorithmic}
\end{algorithm}
\end{minipage}
\hfill
\begin{minipage}[!t]{0.448\textwidth}
\begin{algorithm}[H]
\caption{Refinement via diffusion \vphantom{Degradation flow training}}
\label{alg:learnddpm}
\textbf{Input}: Data, $p(\bmsymbol{x})$, trained flow $\bmsymbol{f}_{\bmsymbol{\xi}}$ \vphantom{Data, $p(\bmsymbol{x})$, degradation function $\mathcal{D}$}\\
\textbf{Parameters}: Initial DDPM, $\bmsymbol{\epsilon}_{\bmsymbol{\theta}}$, $\{\beta_t\}_{t=1}^T$ \vphantom{ Initial flow, $\bmsymbol{f}_{\bmsymbol{\xi}}$, $\sigma>0$}\\
\textbf{Output}: Learned $\bmsymbol{\theta}$\vphantom{$\xi$}
\begin{algorithmic}[1] 
\REPEAT
\STATE $[\bmsymbol{y}, \bmsymbol{z}]\sim (\bmsymbol{f}_{\bmsymbol{\xi}})_{\#}p$ \vphantom{$\bmsymbol{x}\sim p$}
\STATE $t\sim\text{Unif}(\{1,\dots,T\})$, $\bmsymbol{\epsilon}\sim\mathcal{N}(\bmsymbol{0},\bmsymbol{I})$ \vphantom{$[\bmsymbol{y}, \bmsymbol{z}]=\bmsymbol{f}_{\bmsymbol{\xi}}(\bmsymbol{x})$ }
\STATE $\bmsymbol{z}_t=\sqrt{\bar\alpha_t}\bmsymbol{z}+\sqrt{1-\bar\alpha_t}\bmsymbol{\epsilon}$
\STATE Take gradient descent step with \\ \centering \vphantom{$\log\left|\det\frac{\partial\left[\bmsymbol{y}, \bmsymbol{z}\right]}{\partial{\bmsymbol{x}}}\right|$}$\nabla_{\bmsymbol{\theta}}\|\bmsymbol{\epsilon}_{\bmsymbol{\theta}}\left(\bmsymbol{z}_t,\bmsymbol{y},t\right)-\bmsymbol{\epsilon}\|^2_2$
\UNTIL{converged}
\RETURN{$\bmsymbol{\theta}$\vphantom{$\xi$}}
\end{algorithmic}\vspace{0.1pt}
\end{algorithm}
\end{minipage}
\subsection{Flow-Based Generative Pseudoinverse}
\label{subsec:flowpinv}
Here we provide a construction of generalized inverses based on the coupling architecture, as discussed in Section \ref{subsec:flows}. Informally, for $\mathcal{D}: \mathcal{X}\to\mathcal{Y}$, we are interested in learning a bijection such that $\mathcal{X}\leftrightarrow\mathcal{Y}\times\mathcal{Z}$, where $\mathcal{Z}$ describes a kind of generalized kernel, which extends from the linear case.
\begin{definition}[Generalized Kernel]The kernel of $\mathcal{D}:\mathcal{X}\to\mathcal{Y}$ at $\bmsymbol{x}$ is, e.g., $\{\bmsymbol{x}'-\bmsymbol{x}:\mathcal{D}(\bmsymbol{x}')=\mathcal{D}(\bmsymbol{x})\}$.
\end{definition}
Intuitively, the kernel is the set of perturbations of solutions to our SR problem such that consistency is preserved. However, it is generally not disentangled from the measurements $\bmsymbol{y}$. To be precise, we learn $\mathcal{X}\leftrightarrow\mathcal{Y}\times\mathcal{Z}$ such that $\mathcal{Z}$ space is regularized to be approximately independent from $\bmsymbol{y}$.

Such transformations may be learned by application of the change-of-variables formula $p_\mathcal{X}(\bmsymbol{x})=p_{\mathcal{Y}}(\bmsymbol{y})p_{\mathcal{Z}|\mathcal{Y}}(\bmsymbol{z}|\bmsymbol{y})\left|\det\frac{\partial[\bmsymbol{y}, \bmsymbol{z}]}{\partial{\bmsymbol{x}}}\right|$. We use this to model $p_{\mathcal{Y}}(\bmsymbol{y})=\mathbb{E}_{\bmsymbol{x}\sim p}\mathcal{N}(\bmsymbol{y};\mathcal{D}(\bmsymbol{x}),\sigma^2\bmsymbol{I})$ with $\sigma\approx0$ and $p_{\mathcal{Z}|\mathcal{Y}}(\bmsymbol{z}|\bmsymbol{y})=\mathcal{N}(\bmsymbol{z};\bmsymbol{0},\bmsymbol{I})$ approximately, in the sense of Algorithm \ref{alg:learnflow}. More generally, for $p_{\mathcal{Y}}(\bmsymbol{y})$, one may model $p_{\mathcal{Y}|\mathcal{X}}(\bmsymbol{y}|\bmsymbol{x}')=\mathcal{N}(\bmsymbol{y};\mathcal{D}(\bmsymbol{x}'),\sigma^2\bmsymbol{I})$\footnote{Since the flow is deterministic, stochastic dequantization of inputs may be used to introduce randomness.} with $\bmsymbol{x}'$ drawn from a chosen $p'(\cdot|\bmsymbol{x})$, as detailed in Appendix \ref{app:vlb}. We summarize by highlighting key properties of our construction below.
\begin{remark}
\label{prop:flowdefines}
Any flow such that $\mathcal{X}\leftrightarrow\mathcal{Y}\times\mathcal{Z}$ defines a degradation process, $\mathcal{D}:\mathcal{X}\to\mathcal{Y}$, along with its reflexive generalized inverses. In particular, elements in $\mathcal{Z}$ characterize the associated kernel.
\end{remark}
\begin{theorem}[Appendix \ref{app:proof}]
\label{prop:asymptoticconsistency}
Suppose $p_{\mathcal{X}}=p$ and $p_{\mathcal{Y}|\mathcal{X}}(\bmsymbol{y}|\bmsymbol{x})=\mathcal{N}(\bmsymbol{y};\mathcal{D}(\bmsymbol{x}),\sigma^2\bmsymbol{I})$. Let $q(\bmsymbol{y}):=\mathbb{E}_{\bmsymbol{x}\sim p}\delta_{\mathcal{D}(\bmsymbol{x})-\bmsymbol{y}}$. We have the consistency guarantee $\mathbb{E}_{\bmsymbol{y}\sim q,\bmsymbol{x}\sim p_{\mathcal{X}|\mathcal{Y}}(\cdot|\bmsymbol{y})}\|\mathcal{D}(\bmsymbol{x})-\bmsymbol{y}\|^2_2\leq2(\dim\mathcal{Y})\sigma^2$.
\end{theorem}
\subsection{Refinement of the Kernel via Diffusion}
Supposing the flow fits the degradation process in $\mathcal{Y}$, any further processing in $\mathcal{Z}$ preserves consistency. To overcome expressivity limitations of normalizing flows, we propose a refinement via DDPMs, shown in Algorithm \ref{alg:learnddpm}. As this happens strictly in $\mathcal{Z}$ space, the DDPM is theoretically unconstrained. Moreover, by designing $\mathcal{Z}$ a priori, to approximate independent and identically distributed Gaussians, we intuitively simplify the reverse diffusion process and facilitate convergence. The complete sampling process with our method is given in Algorithm \ref{alg:sampling}.

\begin{minipage}[b]{0.495\textwidth}
\begin{algorithm}[H]
\caption{SR with generative pseudoinverse}
\label{alg:sampling}
\textbf{Input}: Low-resolution samples $\bmsymbol{y}\sim q$\\
\textbf{Parameters}: Trained $\bmsymbol{f}_{\bmsymbol{\xi}}$, $\bmsymbol{\epsilon}_{\bmsymbol{\theta}}$, $\{\beta_t, \sigma_t\}_{t=1}^T$ \\
\textbf{Output}:~$\bmsymbol{x}\sim p_{\mathcal{X}|\mathcal{Y}}(\cdot|\bmsymbol{y})$
\begin{algorithmic}[1] 
\STATE $\bmsymbol{z}_T\sim\mathcal{N}(\bmsymbol{0},\bmsymbol{I})$
\FOR{$t=T,\dots,1$}
\STATE $\bmsymbol{\mu}_{t-1|t}=\frac{1}{\sqrt{\alpha_t}}\left[\bmsymbol{z}_t-\frac{1-\alpha_t}{\sqrt{1-\bar\alpha_t}}\bmsymbol{\epsilon}_{\bmsymbol{\theta}}(\bmsymbol{z}_t,\bmsymbol{y},t)\right]$
\STATE $\bmsymbol{z}_{t-1}\sim\mathcal{N}(\bmsymbol{\mu}_{t-1|t},\sigma^2_t\bmsymbol{I})$
\ENDFOR
\RETURN{$\bmsymbol{f}^{-1}_{\bmsymbol{\xi}}(\bmsymbol{y},\bmsymbol{z}_0)$}
\end{algorithmic}
\end{algorithm}
\end{minipage}
\hfill
\begin{minipage}[b]{0.495\textwidth}
\begin{table}[H]
    \centering
    \caption{$8\times$ SR methods on CelebA-HQ.}
    \setlength{\tabcolsep}{1mm}
    \begin{tabular}{@{}lrrr@{}}
        \toprule
        Method & PSNR $\uparrow$ & SSIM $\uparrow$ & Consistency $\downarrow$ \\
        \midrule
        PULSE & 16.88 & 0.44 & 161.1 \\
        FSRGAN & 23.01 & 0.62 & 33.8 \\
        SR3-R & 23.96 & 0.69 & 2.71 \\
        SR3-D & 23.04 & 0.65 & 2.68  \\
        IDM & 24.01 & \textbf{0.71} & 2.14 \\
        \textbf{Our method} & & & \\
        7+1 NFEs & \textbf{24.09} & \textbf{0.71} & 0.31 \\
        100+1 NFEs & 23.16
 & 0.67 & \textbf{0.06} \\
        \bottomrule
    \end{tabular}
    \label{tab:facesr}
\end{table}
\end{minipage}
\section{Experiments}
\label{sec:experiments}
We train our system on bicubic degradations for $16\times16 \to 128\times128$ face SR. We use the FFHQ dataset \citep{ffhq} during development and the CelebA-HQ dataset \citep{celebahq} for evaluation. PSNR, SSIM and consistency errors are used to quantify performance. Consistency is calculated as the MSE ($\times10^{-5}$) between the measurements and the downsampled SR images.

For modeling the flow, we use the affine coupling architecture from \citet{irn}. In particular, a single-scale flow is chosen as this greatly simplifies our framework. It is adapted via nearest upsampling, which is trivially invertible, matching lower-resolution measurements to the $\mathcal{Y}$ space. Our DDPM is based on the architecture from \citet{improveddiffusion}, with a linear noise schedule that may be subsampled during evaluation for faster inference.

All experiments are conducted on a Linux cluster with NVIDIA L40S GPUs. The flow is trained for 500k iterations and our DDPM for 1M iterations with batch sizes 16 and 256 respectively. 
In total, our system has 110M parameters, with 31M allocated to the flow and 79M to the DDPM.
\subsection{Results}
We present results for face SR in Figures \ref{fig:facesrexample}, \ref{fig:facesr} and Table \ref{tab:facesr}. We compare our method with PULSE \citep{pulse}, FSRGAN \citep{fsrgan}, SR3 \citep{sr3}, IDM \citep{idm} and DiWa \citep{diwa}. Our algorithm shows strong performance, surpassing IDM in terms of PSNR and consistency with only seven Neural Function Evaluations (NFEs). However, focusing on the last two rows of Table \ref{tab:facesr}, we observe a trade-off between consistency and the other objective metrics. Still, when the NFEs are increased to 100, matching SR3 diffusion, we maintain competitive scores while achieving a further reduction in consistency errors by a factor of five.
\section{Discussion}
Having presented consistency from the perspective of trustworthiness, we expect this research to be valuable in fields such as medical, forensic imaging, remote sensing and autonomous systems, particularly in cases where noise is negligible or has been effectively mitigated. In these scenarios, maintaining the integrity of image reconstructions is paramount as inconsistencies can lead to misinterpretations with significant real-world consequences.

However, there is also an inherent trade-off between minimizing hallucinations and maximizing quality \citep{cohen2024looks}. Therefore, depending on the application, strict consistency may not be ideal and alternatives that instead loosely follow the measurements may be preferred.

We conclude by noting that, although we specifically studied SR, our methods could potentially be generalized beyond inverse problems to arbitrary conditioning mechanisms, enabling deep learning systems to reliably follow guiding signals that may extend to alternative modalities.

\raggedbottom
\bibliography{iclr2025_delta}
\bibliographystyle{iclr2025_delta}

\newpage

\appendix

\section{Variational Lower Bound}
\label{app:vlb}

Let $p'=p'(\cdot|\bmsymbol{x})$ be an arbitrary
variational distribution. By Jensen's inequality, we have
\begin{equation}
\log p_{\mathcal{Y}}(\bmsymbol{y})=\log\mathbb{E}_{\bmsymbol{x}'\sim p}
\mathcal{N}(\bmsymbol{y};\mathcal{D}(\bmsymbol{x}'),\sigma^2\bmsymbol{I})\geq
\mathbb{E}_{\bmsymbol{x}'\sim p'(\cdot|\bmsymbol{x})}
\log\mathcal{N}(\bmsymbol{y};
\mathcal{D}(\bmsymbol{x}'),\sigma^2\bmsymbol{I})-D_{\mathrm{KL}}\bigl(
p'\|p
\bigr).
\end{equation}
Substituting into the change-of-variables formula for $[\bmsymbol{y},\bmsymbol{z}]=\bmsymbol{f}_{\bmsymbol{\xi}}(\bmsymbol{x})$ and taking the expectation gives
\begin{equation}
\begin{split}
\mathbb{E}_{\bmsymbol{x}\sim p}\log p_{\mathcal{X}}(\bmsymbol{x})\geq\mathbb{E}_{\bmsymbol{x}\sim p}\biggl[&\mathbb{E}_{\bmsymbol{x}'\sim p'(\cdot|\bmsymbol{x})}
\log\mathcal{N}(\bmsymbol{y};
\mathcal{D}(\bmsymbol{x}'),\sigma^2\bmsymbol{I})\\&-D_{\mathrm{KL}}\bigl(
p'\|p
\bigr)
+
\log\mathcal{N}(\bmsymbol{z};\bmsymbol{0},\bmsymbol{I})+
\log\left|\det
\frac{\partial[\bmsymbol{y},\bmsymbol{z}]}
{\partial\bmsymbol{x}}\right|\biggr].
\end{split}
\end{equation}
Under this framework, Algorithm \ref{alg:learnflow} may be intuitively recovered by taking $p'(\bmsymbol{x}'|\bmsymbol{x})\approx\delta_{\bmsymbol{x}'-\bmsymbol{x}}$ and renormalizing away the divergence, which becomes independent of the flow parameters anyway.
\section{Theorem \ref{prop:asymptoticconsistency}}
\label{app:proof}

\begin{proof}
In fact, we can demonstrate a sharper result. Write $d=\dim\mathcal{Y}$ and, for $\tau>0$, define
\begin{equation}
q_\tau(\bmsymbol{y})
=\mathbb{E}_{\bmsymbol{x}\sim p}
\mathcal{N}(\bmsymbol{y};\mathcal{D}(\bmsymbol{x}),\tau^2\bmsymbol{I}).
\end{equation}
The assumptions imply $p_{\mathcal{Y}}=q_\sigma$. Hence, by Jensen's inequality and Bayes' rule,
\begin{equation}
\begin{aligned}
\mathbb{E}_{\bmsymbol{x}\sim p_{\mathcal{X}|\mathcal{Y}}(\cdot|\bmsymbol{y})}
\|\mathcal{D}(\bmsymbol{x})-\bmsymbol{y}\|_2^2&\leq
4\sigma^2\log
\mathbb{E}_{\bmsymbol{x}\sim p_{\mathcal{X}|\mathcal{Y}}(\cdot|\bmsymbol{y})}
\exp\left(\frac{\|\mathcal{D}(\bmsymbol{x})-\bmsymbol{y}\|_2^2}{4\sigma^2}\right)\\
&=
4\sigma^2\left(
\frac{d}{2}\log 2+
\log\frac{q_{\sqrt{2}\sigma}(\bmsymbol{y})}{q_\sigma(\bmsymbol{y})}
\right).
\end{aligned}
\end{equation}
It remains to bound the expected log density ratio. Let
$\bmsymbol{\epsilon}\sim\mathcal{N}(\bmsymbol{0},\frac{\sigma^2}{2}\bmsymbol{I})$,
independently of $\bmsymbol{y}$. Gaussian convolution and Jensen's inequality give the lower bound
\begin{equation}
\log q_\sigma(\bmsymbol{y})
=\log\mathbb{E}_{\bmsymbol{\epsilon}}
q_{\sigma/\sqrt{2}}(\bmsymbol{y}+\bmsymbol{\epsilon})
\geq
\mathbb{E}_{\bmsymbol{\epsilon}}
\log q_{\sigma/\sqrt{2}}(\bmsymbol{y}+\bmsymbol{\epsilon}).
\end{equation}
Next, we focus on upper bounding $\log q_{\sqrt{2}\sigma}(\bmsymbol{y})$. For fixed $\bmsymbol{y}$, define the probability density
\begin{equation}
r_{\bmsymbol{y}}(\bmsymbol{x})
=
\frac{p(\bmsymbol{x})
\mathcal{N}(\bmsymbol{y};\mathcal{D}(\bmsymbol{x}),2\sigma^2\bmsymbol{I})}
{q_{\sqrt{2}\sigma}(\bmsymbol{y})}.
\end{equation}
Expanding the Gaussian exponent gives
\begin{equation}
\frac{q_{\sqrt{2}\sigma}(\bmsymbol{y}+\bmsymbol{\epsilon})}
{q_{\sqrt{2}\sigma}(\bmsymbol{y})}
=
\mathbb{E}_{\bmsymbol{x}\sim r_{\bmsymbol{y}}}
\exp\left(
\frac{\bmsymbol{\epsilon}^{\top}
(\mathcal{D}(\bmsymbol{x})-\bmsymbol{y})}{2\sigma^2}
-\frac{\|\bmsymbol{\epsilon}\|_2^2}{4\sigma^2}
\right).
\end{equation}
Applying Jensen's inequality we obtain
\begin{equation}
\log\frac{q_{\sqrt{2}\sigma}(\bmsymbol{y}+\bmsymbol{\epsilon})}
{q_{\sqrt{2}\sigma}(\bmsymbol{y})}
\geq
\frac{\bmsymbol{\epsilon}^{\top}}{2\sigma^2}
\mathbb{E}_{\bmsymbol{x}\sim r_{\bmsymbol{y}}}
[\mathcal{D}(\bmsymbol{x})-\bmsymbol{y}]
-\frac{\|\bmsymbol{\epsilon}\|_2^2}{4\sigma^2}.
\end{equation}
Averaging over $\bmsymbol{\epsilon}\sim
\mathcal{N}(\bmsymbol{0},\frac{\sigma^2}{2}\bmsymbol{I})$ and rearranging yields
\begin{equation}
\log q_{\sqrt{2}\sigma}(\bmsymbol{y})
\leq
\mathbb{E}_{\bmsymbol{\epsilon}}
\log q_{\sqrt{2}\sigma}(\bmsymbol{y}+\bmsymbol{\epsilon})
+\frac{d}{8}.
\end{equation}
Now, subtracting the point-wise inequalities for $\log q_{\sqrt{2}\sigma}$, $\log q_{\sigma}$ and using
$\bmsymbol{y}+\bmsymbol{\epsilon}\sim q_{\sigma/\sqrt{2}}$ for $\bmsymbol{y}\sim q$
\begin{equation}
\mathbb{E}_{\bmsymbol{y}\sim q}
\log\frac{q_{\sqrt{2}\sigma}(\bmsymbol{y})}{q_\sigma(\bmsymbol{y})}\leq \frac{d}{8}
+\mathbb{E}_{\bmsymbol{y}\sim q_{\sigma/\sqrt{2}}}
\log\frac{q_{\sqrt{2}\sigma}(\bmsymbol{y})}
{q_{\sigma/\sqrt{2}}(\bmsymbol{y})}\leq \frac{d}{8},
\end{equation}
where the last inequality follows from a final application of Jensen, since densities integrate to one.

Averaging the first bound over $\bmsymbol{y}\sim q$ and invoking this last upper bound gives
\begin{equation}
\mathbb{E}_{\bmsymbol{y}\sim q,\,
\bmsymbol{x}\sim p_{\mathcal{X}|\mathcal{Y}}(\cdot|\bmsymbol{y})}
\|\mathcal{D}(\bmsymbol{x})-\bmsymbol{y}\|_2^2\leq
\left(2\log 2+\frac{1}{2}\right)d\sigma^2
\leq 2(\dim\mathcal{Y})\sigma^2.
\end{equation}
\end{proof}

\end{document}